# Strong Secrecy and Reliable Byzantine Detection in the Presence of an Untrusted Relay

Xiang He   Aylin Yener

Wireless Communications and Networking Laboratory

Electrical Engineering Department

The Pennsylvania State University, University Park, PA 16802

*xxh119@psu.edu   yener@ee.psu.edu*

March 31, 2010

## Abstract

We consider a Gaussian two-hop network where the source and the destination can communicate only via a relay node who is both an eavesdropper and a Byzantine adversary. Both the source and the destination nodes are allowed to transmit, and the relay receives a superposition of their transmitted signals. We propose a new coding scheme that satisfies two requirements simultaneously: the transmitted message must be kept secret from the relay node, and the destination must be able to detect any Byzantine attack that the relay node might launch reliably and fast. The three main components of the scheme are the nested lattice code, the privacy amplification and the algebraic manipulation detection (AMD) code. Specifically, for the Gaussian two-hop network, we show that lattice coding can successfully pair with AMD codes enabling its first application to a noisy channel model. We prove, using this new coding scheme, that the probability that the Byzantine attack goes undetected decreases exponentially fast with respect to the number of channel uses, while the loss in the secrecy rate, compared to the rate achievable when the relay is honest, can be made arbitrarily small. In addition, in contrast with prior work in Gaussian channels, the notion of secrecy provided here is strong secrecy.

This work was presented in part at International Symposium on Information Theory, ISIT 2009, July, 2009, and the Information Theory Workshop, ITW 2010, January 2010. This work is supported in part by the National Science Foundation with Grant CNS-0716325, and the DARPA ITMANET Program with Grant W911NF-07-1-0028.

## I. INTRODUCTION

Information theoretic secrecy, first proposed by Shannon [1], provides confidentiality of transmitted information against an adversary regardless of its computational power. Shannon proved that if the adversary has access to the signals transmitted by the sender of the secret message through a noiseless channel, then, to achieve perfect secrecy from the adversary, the sender and the receiver has to share a secret key of the same length as the message. Although Shannon's result implied that secret communication was impractical in this setting, it was later shown by Wyner [2] that this pessimistic result was a consequence of the noiseless channel assumption. Specifically, it was shown that when the adversary has noisy observations of the signals transmitted by the sender, a nonzero transmission rate for the secrecy message is achievable without requiring the transmitter to pre-share a key with the receiver [2]–[4]. More recently, the fundamental rate limits at which the secret communication can take place in the presence of an eavesdropper were studied for a number of multi-terminal models, e.g., the broadcast channel [5], [6], the two-way channel [7], [8], the multiple access channel [7] and the interference channel [9], [10].

Secure communication for channel models with a relay node has been studied from a variety of perspectives, including the relay node as a helper to the legitimate communication link [11], or to an eavesdropper [12]. References [13]–[16] consider the case where the relay node itself is the eavesdropper from whom the information transmitted from the source to the destination must be kept secret. This setting, which provides theoretical foundations toward the utilization of untrusted relay nodes in network design, is relevant in practice: The potentially untrusted routers of today's Internet routinely relay sensitive information for its users. The current approach is that the authenticity and secrecy of the information is protected by security protocols assuming these routers are *limited in computational power* [17]. It is interesting to address the role of these routers if they are computational power unlimited adversaries.

To answer this question, in [8], [14], [15], as a first step, we considered the case where the relay node was "honest but curious". This means that the curious relay node is not trusted with confidential messages. On the other hand, it is honest, and thus conforms to the system rules and performs the designated relaying scheme. Reference [14] considered the three-node relay network with such a relay. References [8], [15] considered the two-way relay channel where two nodes could only communicate through such a relay node. In these works, we showed that if the relay was not trusted but honest, recruiting it to help relay information was *useful* in achieving a higher secrecy rate than simply treating the relay node as an eavesdropper. This effect is most pronounced in the two-hop model studied in [15], in which the achievable rate is $0$ if the relay node is excluded from communication, and increases to being within 1bit of the rate of having trusted relay if the untrusted relay node is properly utilized. Similar observations can be made in networks with multiple confidential messages [16].

It is the next natural step to consider the problem where the relay node is curious and is potentially *dishonest*. This means that the relay can deviate from its designated behavior. This can be as benign as the relay node experiencing a failure and stopping transmission, which is obviously easy to detect. However, if the relay is a malicious entity (or is captured by one), a more detrimental scenario can materialize. Specifically, the relay can attempt to deceive the destination into accepting a counterfeit message by actively manipulating the signals it relays. Such behavior is a "Byzantine attack" [18]. When the adversary is limited in computational power, this type of attack can be detected via message authentication code or digital signatures [17]. The security guarantee promised by these schemes is essentially based on the absence of known effective attack strategies and the fact that their reliability can be proved if a very small set of assumptions is made.

In this work, we tackle the case where the Byzantine adversary has unlimited computational power. In an effort to demonstrate the simplest network which relies on an untrusted node to communicate, we consider a two-hop network [15]. In contrast to reference [15], which considered an honest but curious relay, we allow the relay node to actively modify the transmitted signal in any way it desires. The goal of the destination thus becomes detecting the message that has been altered fast and reliably whenever the relay node chooses to do so.

Toward accomplishing this goal, there are several known results that can be leveraged, each with their own limitations. For example, Byzantine attack detection can be viewed as an authentication problem, by treating the counterfeit message $W'$ as a message from a "wrong" source node. An information theoretic secrecy scheme with an authentication capability was proposed in [19]. However, like other message authentication codes [20], the source has to share an authentication key with the destination beforehand.

It is known, on the other hand, that to detect the Byzantine attack, which is a milder require-ment than authentication, it is not essential to share keys. In reference [21], the so-called algebraic manipulation detection (AMD) code was used for encoding the data from the source node which ensures the probability that the Byzantine attack succeeds can be made arbitrarily small with an arbitrarily small loss in rate. A limitation of this scheme is that it has to be used along with a secrecy sharing scheme that has certain *linearity* property [21], which is easily fulfilled in a noiseless network as shown in [18], [22]. Indeed, in [22], we considered a deterministic two-hop network and it was shown that by using AMD code, the probability that the Byzantine adversary wins decreases exponentially fast with respect to the total number of channel uses $n'$ while the loss in rate can be made arbitrarily small. On the other hand, for *noisy* channels, secret sharing schemes generally fail to have the required linearity property. As a result, to date the strongest result that could have been obtained is that, for a noisy two-hop network, the probability that a Byzantine attack goes undetected decreases exponentially only with respect to $\sqrt{n'}$ in [22].

The main contribution of this work is to demonstrate that for the Gaussian two-hop network, the probability that a Byzantine attack goes undetected, i.e., the adversary wins, also decreases exponentially fast with respect to $n'$, while the loss in secrecy rate can be made arbitrarily small. Hence, the same result achievable for the deterministic two-hop network is attainable for this *noisy* two-hop network. This represents a departure from traditional security approaches that assume a noiseless bit pipe for communication and brings the physical characteristics of the channel into the picture while providing a guarantee thought to be possible only with the noiseless setting. The key to prove this result is the introduction of a new strong secrecy scheme. Its existence is proved via the representation theorem derived in [10], [23] and the privacy amplification technique presented in [24], [25]. Compared to previously known strong secrecy schemes, the main differences are:

1) Unlike the randomly generated codes in [26], the decoder of the new scheme is linear for certain rate configurations.
2) Unlike [10], [23], the codeword consists of a single lattice point rather than multiple lattice points. This allows the mutual information between the confidential message and eavesdropper's observation to decrease exponentially with respect to $n'$. Hence the notion of secrecy provided by this scheme is stronger than commonly used strong secrecy scheme, which only requires this mutual information to vanish with respect to $n'$.

The first item provides the linear property required by AMD code. The stronger-than-usual secrecy notion in the second item is essential in preserving the Byzantine detection performance offered by AMD code. As will be shown in Section VI, the commonly used strong secrecy notion, as in [25], [27], is insufficient for this purpose.

There is other work in Byzantine detection from which this work differs. Notably, reference [28] proposed to use the sender of the confidential message to monitor the behavior of the relay node. This so-called "watchdog" scheme could also have been used in the setting we consider if the message in transmission were not to be kept secret from the relay node. However, when the message is confidential, using a "watchdog" is not possible. This is because there is no direct link between the two legitimate communicating nodes which means the sender has no information regarding the signals transmitted by the destination. As will be explained in Section IV, these signals are necessary in order to deploy cooperative jamming [7] to keep the message secret from the relay node, see also [15]. Since the received signals at the relay is garbled by signals transmitted by the destination, so are the signals transmitted from it. This prevents the source from detecting whether the relay misbehaves by just looking at its transmitted signals without the knowledge of the signals transmitted from the destination.

This work should also be differentiated from references [29]–[32]. In these works, the adversaries can also actively manipulate the signals received by the destination. However, the purpose is to find a way for reliable communication in the presence of such adversaries carrying out the worst-case attack. In the two-hop network considered in this work, this is not possible since there is no direct link between the two legitimate communicating nodes. Hence, when Byzantine behavior is detected, we need to forgo the relay.

The remainder of the paper is organized as follows: In Section II, we describe the system model and formulate the Byzantine detection and secrecy problem. In Section III, we review known Byzantine detection schemes, in particular, the AMD code and describe the technical obstacles to be overcome in this work. Section IV-VI describe the main components of strongly secure scheme proposed in this work and how it can be combined with AMD codes for Byzantine detection purpose. Section VII concludes the paper.

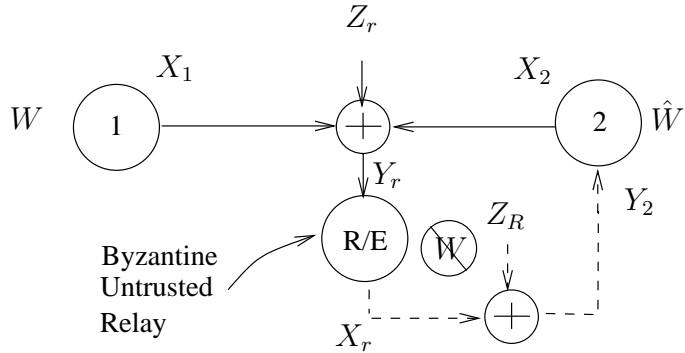

Fig. 1. The Gaussian two-hop network. Phase 1 is indicated by solid line, and phase 2 by dashed line. R/E: Relay/Eavesdropper. $Y_1$ is not shown.

## II. SYSTEM MODEL AND PROBLEM FORMULATION

The Gaussian two-hop network with a Byzantine relay node is shown in Figure 1. In this model, node 1 wants to send a confidential message $W$ to node 2. Since it can not communicate with node 2 directly, it recruits the help of a relay node, who is not trusted with the message $W$. The signal received by the relay node consists of the signals transmitted by both node 1 and 2, and the signal broadcasted by the relay node is heard by both nodes as well. These are fitting assumptions for wireless communication. Let $X_i, i = 1, 2, X_r$ denote the signal transmitted by node 1, 2 and the relay. Let $Y_i, i = 1, 2$ and $Y_r$ denote their received signals respectively. After normalizing the channel gains, we have

$$Y_r = X_1 + X_2 + Z_r \tag{1}$$

$$Y_2 = X_r + Z_R, \quad Y_1 = hX_r + Z'_R \tag{2}$$

where $Z_r$, $Z_R$ and $Z'_R$ are independent Gaussian random variables with zero mean and unit variance. $h$ is the normalized channel gain. Since $Y_1$ is not used in the scheme described in this work, it is omitted in Figure 1 for clarity. We assume each node is half-duplex. For simplicity, we assume the relay node transmits during half of all channel uses. Without loss of generality, we assume node 1 and 2 do not transmit when the relay node transmits since the relay node can not receive and relay their transmitted signals simultaneously. We also assume during the $n$ channel uses that the relay node transmits, its transmission power averaged over these channel uses should not exceed $\bar{P}$. During the remaining $n$ channel uses that node 1 and 2 may transmit, the transmission power of each of these two nodes averaged over these channel uses should not exceed $\bar{P}$.

We assume the Byzantine adversary at the relay node can employ any stochastic function to compute its current transmitted signal. Let $X_{r,i}$ be its transmitted signal at the $i$th channel use. Let $M_r$ be the local randomness available to the relay node. Let $Y_r^{i-1}$ be the signals it received in the past. Let $W$ be the confidential message it is currently relaying. Let $f_i$ be the relaying function. Then the attacker (relay) can compute:

$$X_{r,i} = f_i(M_r, Y_r^{i-1}, W) \qquad (3)$$

It might seem inconsistent at first glance to assume the Byzantine adversary knows the message, which should be kept secret from the relay node in the first place. However, when the possible choice for $W$ are limited, for example, to being binary, the attacker has a non-negligible probability of success for guessing it. This can also happen when the channel is used to transmit data with high redundancy and stringent latency requirement, so that adjacent messages are highly likely to share the same value. If, somehow, the adversary has access to earlier messages, it can guess the value of the current message with high probability of success. As a result, it is a common practice to design a reliable message authentication scheme by assuming the adversary knows the message [20, Definition 4.2]. Here too, we follow this convention.

The Byzantine detection problem for secure communication using an untrusted relay can be stated as follows:

Let the total number of channel uses be $n' = 2n$, during which each node transmits during $n$ channel uses. Let $\hat{W}$ be the estimate of $W$ computed by the destination, i.e., node 2, based on its observation. Note that because the relay can be a Byzantine adversary, node 2 may or may not accept $\hat{W}$ as a genuine message from node 1 based on certain criteria.

*Definition 1:* [20] A function of $n$, $\gamma_n$ is *negligible* if for any polynomial of $n$ with a finite degree $\text{poly}(n)$, we have:

$$\lim_{n \to \infty} \text{poly}(n) \gamma_n = 0 \qquad (4)$$

☐

We wish to find the secrecy rate $R_e$ of $W$, defined as

$$R_e = \lim_{n \to \infty} \frac{1}{n'} H(W) \qquad (5)$$

such that the following conditions hold:

1) When the relay node is honest, and $W$ is uniformly distributed over the message set, then both $\Pr\left(W \neq \hat{W}\right)$ and

$$\Pr\left(\hat{W} \text{ is not accepted by Node 2} | W = \hat{W}\right) \tag{6}$$

should be negligible as per Definition 1. Hence, the transmission of $W$ is *reliable*.

2) For $\forall w_0$ in the message set, the probability that the adversary wins, $\Pr(A \text{ } wins)$, given by

$$\Pr(A \text{ } wins) = \Pr\left(\hat{W} \text{ is accepted by Node 2} | W = w_0, W \neq \hat{W}\right) \tag{7}$$

is negligible. Hence any modification on $W$ is detected reliably.

3) $I\left(W; Y_r^n\right)$ is negligible. Since $Y_r^n$ is the observation of the eavesdropper, this means the information that the adversary has regarding the value of $W$ is negligible.

*Remark 1:* Observe that the condition of reliable Byzantine detection in 2) is independent from the distribution of $W$. □

### III. KNOWN BYZANTINE DETECTION SCHEMES

As mentioned in the introduction, when there are no secrecy concerns at the relay, whether the relay is honest or not can be checked by the source node, i.e., node 1, by examining $Y_1$. However, since there are secrecy constraints in our model, applying sender-based Byzantine detection approach is not feasible. Therefore, we will concentrate on a receiver-based approach called algebraic manipulation detection (AMD) code in the sequel.

AMD code was formally defined in [21]. An AMD codeword is composed of three parts: $\{s, x, h\}$, where $s$ is the $d \times 1$ vector on $\mathcal{GF}(q^r)$ representing the message. The component $x$ is called the random seed and is generated from $\mathcal{GF}(q^r)$ by the encoder itself. $h$ is the hash tag and is computed according to the *hash rule*:

$$h = x^{d+2} + \sum_{i=1}^{d} s_i x^i \tag{8}$$

where $s_i$ is the $i$th component of $s$ and the addition and multiplication is defined over $\mathcal{GF}(q^r)$. Suppose the node 2 receives $s', x', h'$, where $s' \neq s$. Let $\Delta_x = x' - x$. $\Delta_h = h' - h$. Then [21] has the following result:

*Theorem 1:* [21, Theorem 2] Assume at least one of $s' - s, \Delta_x, \Delta_h$ is not zero. If the distribution of $x$ conditioned on $\{\Delta_x, \Delta_h, s', s\}$ is uniform over the field $\mathcal{GF}(q^r)$, $q$ being a prime, and $d+2$ is not divisible by $q$, then the probability that the hash rule (8) holds for $\{s', x', h'\}$ is bounded by $\frac{d+1}{q^r}$.

*Remark 2:* The rate of the AMD code is $\frac{d}{d+2}$. The rate can be made arbitrarily close to 1 by choosing a large enough value for $d$.

On the other hand, an AMD codeword can be represented by less than $(d+2)r\log_2 q + 1$ bits. Hence, if we fix $d$ and $q$, the codeword length is a linear function of $r$. Consequently, for a given code rate, the probability that $\{s', x', h'\}$ can pass the hash rule check (8) decreases exponentially fast with respect to the codeword length. $\square$

Despite the excellent performance of the AMD code, applying it in a noisy channel is difficult. This is exemplified by the condition in Theorem 1: The distribution of $x$ conditioned on $\{\Delta_x, \Delta_h, s', s\}$ must be uniform over the field $\mathcal{GF}(q^r)$. In a noisy channel, in general, $\Delta_x$ and $x$ are not independent. In the two-hop network considered in this work, this can be seen from the expression of $\Delta_x$. Let $g$ be the decoding function used by node 2. Let $Y_2^n$ be the signal received by node 2 if relay is honest. Otherwise, we denote it with $\tilde{Y}_2^n$. Assuming the decoding result is correct at all nodes if the relay is honest. In this case, $\Delta_x$ is given by:

$$\Delta_x = x' - x \tag{9}$$
$$= g\left(\tilde{Y}_2^n, X_2^n\right) - g\left(Y_2^n, X_2^n\right) \tag{10}$$

By observing (10), we notice the condition in Theorem 1 can be fulfilled if $g$ is linear in its first parameter and $\tilde{Y}_2^n - Y_2^n$ is independent from $x$. In general, $g$ is not linear. Even if this is the case, it is also difficult to achieve independence between $\tilde{Y}_2^n - Y_2^n$ and $x$. Since both $\tilde{Y}_2^n$ and $Y_2^n$ are signals transmitted by the relay corrupted by the channel noise, the joint distribution of $\tilde{Y}_2^n - Y_2^n$ and $x$ can be made close to an independent distribution if the relay node has negligible information regarding the value of $x$. But it remains to see whether the performance guarantee in Theorem 1 can be preserved when $\tilde{Y}_2^n - Y_2^n$ and $x$ are almost independent rather than truly independent. In the sequel, we will propose a strong secrecy scheme that overcomes these problems.

## IV. Lattice Coding Scheme

We first briefly review the communication scheme when the relay is "honest but curious", on top of which we will build the strong secrecy scheme and the Byzantine detection scheme in the sequel.

Since each node is half-duplex, naturally we have a two-phase scheme. In phase one, nodes 1 and 2 transmit, and the relay node receives. In phase two, the relay transmits. For simplicity, we assume that each phase occupies the same number of channel uses. It was shown in [15] that these two phases can be used to facilitate the transmission of the confidential message $W$ from node 1 to 2: The channel alternates between phase one and phase two. During phase one, node 1 transmits the confidential message via $X_1$ and at the same time node 2 sends a signal $X_2$ to jam the relay node. During phase two, the relay node transmits to node 2 based on the signal it received during phase one. Since node 2 knows $X_2$, it can subtract it to obtain a clean signal. The relay node, however, does not know $X_2$ and hence can only observe a noisy version of $X_1$. Intuitively, this means node 1 can transmit to node 2 at a rate higher than the relay node can decode, and that this excess rate can be used to convey confidential messages. This idea was formalized in [15] using compress-and-forward relaying and in [23] using compute-and-forward relaying. In this work, we focus on the compute-and-forward scheme as it offers the algebraic structure that facilitates detection of a Byzantine attack.

In the compute-and-forward scheme, the signals transmitted by the two legitimate nodes are taken from the same nested lattice codebook. This scheme was first proposed in [33] for a Gaussian two-way relay channel without eavesdroppers. Later, the scheme was used in [23] as a building block to transmit confidential messages when the relay is honest but curious, i.e., is an eavesdropper but not a Byzantine adversary. The lattice coding scheme is described next for completeness:

We begin by introducing basic notations for the nested lattice structure: For a lattice $\Lambda_c$, the modulus operation $x \mod \Lambda_c$ is defined as $x \mod \Lambda_c = x - \arg\min_{t \in \Lambda_c} d(x, t)$, where $d(x, t)$ is the Euclidean distance between $x$ and $t$. The fundamental region of a lattice $\mathcal{V}(\Lambda_c)$ is defined as the set $\{x : x \mod \Lambda_c = x\}$. A pair of $N$-dimensional lattices $\{\Lambda, \Lambda_c\}$ is said to have a nested structure if $\Lambda_c \subset \Lambda$ [34].

Now consider a pair of $N$-dimensional nested lattice pair $\{\Lambda, \Lambda_c\}$ which is properly designed as in [34]. The signal transmitted by each node is given by

$$X_i^N = \left(t_i^N + d_i^N\right) \mod \Lambda_c, \quad i = 1, 2 \tag{11}$$

where $t_i^N \in \Lambda \cap \mathcal{V}(\Lambda_c)$, and $d_i^N, i = 1, 2$ are two fixed vectors in $\mathcal{V}(\Lambda_c)$ and are known by the relay node. For our purpose, $t_1^N$ will be computed from the confidential message. $t_2^N$ is independent from $t_1^N$ and is chosen from $\Lambda \cap \mathcal{V}(\Lambda_c)$ according to a uniform distribution. As a result, $X_2^N = t_2^N + d_2^N \mod \Lambda_c$ serves as the jamming signal to confuse the untrusted relay node.

An honest relay node will then decode $t_1^N + t_2^N \mod \Lambda_c$ and transmit $t_1^N + t_2^N + d_3^N \mod \Lambda_c$ during phase two, where $d_3^N$ is a fixed vector in $\mathcal{V}(\Lambda_c)$ and is known by node 2. Node 2 then decodes $\hat{t}^N = t_1^N + t_2^N \mod \Lambda_c$ from the signal it received during phase two. An estimate of $t_1^N$, denoted by $\hat{t}_1^N$, is then by computed from $\hat{t}^N - t_2^N \mod \Lambda_c$.

Define $|\mathcal{S}|$ be the cardinality of a set $\mathcal{S}$. Define $R_0$ as

$$R_0 = \frac{1}{N} \log_2 |\Lambda \cap \mathcal{V}(\Lambda_c)| \tag{12}$$

Then it was shown in [33] that, if

$$R_0 < \frac{1}{2} \log_2(\frac{1}{2} + P) \tag{13}$$

the probability $\Pr(\hat{t}_1^N \neq t_1^N)$ decreases exponentially with respect to $N$.

*Remark 3:* It is clear that if the relay chooses to transmit $t_3^N + d_3^N \mod \Lambda_c$ for some arbitrary $t_3^N \in \Lambda \cap \mathcal{V}(\Lambda_c)$, then node 2 will be forced to accept a message that is not originated from node 1. This shows that unless some proper measure is taken, Byzantine attack can quite easily succeed in this scenario. □

*Remark 4:* $d_i^N, i = 1, 2, 3$ are conventionally defined as random variables uniformly distributed over $\mathcal{V}(\Lambda_c)$ [34]. The reason of defining them to be random is that it is easier to analyze the average error performance of an ensemble of lattice code books parameterized by the dithering vectors than to analyze the error performance of a specific lattice code book [35]. However, from the result on the average performance, we can also claim that there must exist some fixed $d_i^N, i = 1, 2, 3$, which corresponds to fixed lattice codebooks in the ensemble, and these $d_i^N, i = 1, 2, 3$ also provide vanishing error probability and meet the average power constraints [10]. Hence in the sequel we assume $d_i^N, i = 1, 2, 3$ are fixed. □

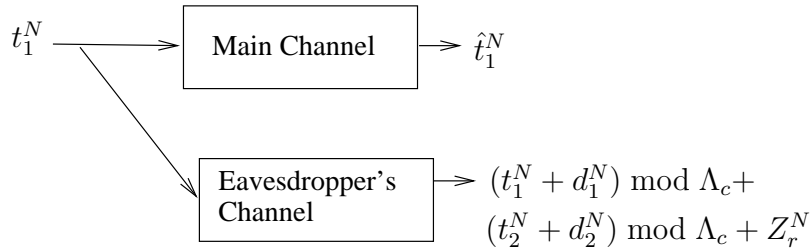

Fig. 2. The lattice input Wiretap Channel

## V. Using Nested Lattice Codes to Provide Strong Secrecy

For the lattice coding scheme described in Section IV, the two-hop network is equivalent to the *lattice input wiretap channel* shown in Figure 2. The main channel takes input $t_1^N \in \Lambda \cap \mathcal{V}(\Lambda_c)$, and produces output $\hat{t}_1^N$. The eavesdropper channel also takes input $t_1^N$, and has the same observation as the signals received by the relay node in the two-hop network. The only difference from the original two-hop network is that in the two-hop network, it takes another $N$ channel uses for the relay to relay the lattice point to node 2 during which node 1 and 2 do not transmit. Here, to simplify the argument, we omit this detail and will take these additional channel uses into account when we revisit the two-hop network in Section VI. Here, we simply assume that in the lattice-input wiretap channel, the transmitter transmits in each channel use and its average power constraint is given by $P$.

In the sequel, we will design a coding scheme for the lattice-input wiretap channel to transmit a confidential message $W$ reliably such that the following strong secrecy condition holds:

$$I(W; Y_r^N) < \exp(-\bar{\alpha} N), \quad \bar{\alpha} > 0 \tag{14}$$

A sufficient condition for (14) to hold is:

$$I(W; \bar{Y}_r^N) < \exp(-\bar{\alpha} N), \quad \bar{\alpha} > 0 \tag{15}$$

where $\bar{Y}_r^N$ is obtained by subtracting the channel noise $Z_r^N$ from $Y_r^N$:

$$\bar{Y}_r^N = (t_1^N + d_1^N) \bmod \Lambda_c + (t_2^N + d_2^N) \bmod \Lambda_c \tag{16}$$

### A. Strongly Secure Scheme

*1) When $\Lambda_c = q\Lambda$ for a prime $q$:* The self-similar nested lattice code with prime nesting ratio, i.e., $\Lambda_c = q\Lambda$, is a special case of the good nested lattice ensemble proposed in [34, Section 7]. We first consider this case since when $q$ is a prime, the set $(\Lambda + d^N) \cap \mathcal{V}(\Lambda_c)$ is isomorphic to a finite field, as shown by the following lemma:

*Lemma 1:* When $\Lambda_c = q\Lambda$ for a prime $q$ and the generation matrix of $\Lambda$ has full rank, $(\Lambda + d^N) \cap \mathcal{V}(\Lambda_c)$, for the modulus-$\Lambda_c$ plus operation, is isomorphic to the group of a finite field $\mathcal{GF}(q^N)$.

*Proof:* The proof is provided in Appendix A. ∎

*Remark 5:* The isomorphism in Lemma 1 is not affected by the choice of $d$. The fixed dithering vector $d$ is simply used to control the average power of the lattice code book. □

As we will show later in the proof of Theorem 2, the isomorphism property proved by Lemma 1 allows the resulting decoder to be linear and proves to be of critical importance in the Byzantine detection scheme in Section VI.

The next theorem declares the existence of the strong secrecy scheme.

*Theorem 2:* For a given constant $\varepsilon > 0$ that can be arbitrarily small, assume $q$ is a prime large enough such that

$$1 - \frac{1+\varepsilon}{\log_2 q} > 0 \tag{17}$$

Then for an integer $r$, such that

$$0 \leq r \leq N\left(1 - \frac{1+\varepsilon}{\log_2 q}\right) \tag{18}$$

there exists a linear mapping $\mathbf{g}$ from $\mathcal{GF}(q)^N$ to $\mathcal{GF}(q)^r$ such that

1) $\mathbf{g}$ has full row rank $r$.
2) When $t_i^N, i = 1, 2$ are uniformly distributed over $(\Lambda + d_i^N) \cap \mathcal{V}(\Lambda_c)$ and are independent of each other, there exists a positive constant $\beta$ such that

$$I\left(\mathbf{g}\left(t_1^N\right); \bar{Y}_r^N\right) \leq 2e^{-\beta N} \tag{19}$$

Before proving the theorem, we need several supporting results:

First, the following representation theorem from [23] is useful:

*Theorem 3:* [23] For any $u_1, u_2$, such that $u_i \in \mathcal{V}(\Lambda_c), i = 1, 2$, $\sum_{k=1}^{2} u_k$ is uniquely determined by $\{T, \sum_{k=1}^{2} u_k \mod \Lambda_c\}$, where $T$ is an integer such that $1 \leq T \leq 2^N$.

Based on Theorem 3, $\bar{Y}_r^N$ in (16) can be represented by $\{(\sum_{i=1}^{2}(t_i^N + d_i^N)) \bmod \Lambda_c, T\}$. Since $d_i^N, i = 1, 2$ are known by each node, this means $\bar{Y}_r^N$ in (16) can be represented by $\{(t_1^N + t_2^N) \bmod \Lambda_c, T\}$.

We also need the following result which says most matrices have full rank:

*Lemma 2:* Let $\mathbf{G}$ be taken from the set of linear mappings from $\mathcal{GF}(q)^N$ to $\mathcal{GF}(q)^r$ according to a uniform distribution. Hence $\mathbf{G}$ can be represented as a matrix over $\mathcal{GF}(q)$ with $r$ rows and $N$ columns. The probability that $\mathbf{G}$ has full row rank is greater than $1 - q^{r-N}$.

*Proof:* Let $g_i, i = 1, ..., r$ be the $i$th row of $\mathbf{G}$. Then $\mathbf{G}$ does not have full row rank if and only if

$$a_1 g_1 + a_2 g_2 + ... + a_r g_r = 0, \quad a_i \in \mathcal{GF}(q) \tag{20}$$

Since at least one $a_i$ has to be non-zero, there are $q^r - 1$ possible choices for $a_i$.

For each choice of $\{a_i\}$, since one $a_i$ is not zero, there are $q^{N(r-1)}$ solutions for $\{g_i\}$. Hence there are at most $q^{N(r-1)}(q^r - 1)$ $\mathbf{G}$s that do not have full row rank. There are $q^{Nr}$ possible $\mathbf{G}$s in all, each chosen with equal probability. Hence the probability that $\mathbf{G}$ does not have full row rank smaller than $q^{r-N}$, and we have Lemma 2. ∎

Finally, we need the following results on privacy amplification [24], which we state here for completeness: We begin with a couple of useful definitions:

*Definition 2:* For a discrete random variable $X$, the Rényi entropy $H_2(X)$ is defined as

$$H_2(X) = -\log_2 \sum_x \Pr(X = x)^2 \tag{21}$$

The Shannon entropy $H(X)$ is defined as

$$H(X) = -\sum_x \Pr(X = x) \log_2 \Pr(X = x) \tag{22}$$

*Definition 3:* [24, Definition 1] A set of functions $\mathcal{A} \to \mathcal{B}$ is a class of *universal hash function* if for a function $g$ taken from the set according to a uniform distribution, and $x_1, x_2 \in \mathcal{A}, x_1 \neq x_2$, the probability that $g(x_1) = g(x_2)$ holds is at most $1/|\mathcal{B}|$.

We next state the results based on these definitions:

*Lemma 3:* [24] The set of linear mapping as defined in Lemma 2 is a class of universal hash function.

*Theorem 4:* [24, Corollary 4] Let **G** be selected according to a uniform distribution from a class of universal hash function from $\mathcal{A}$ to $\mathcal{GF}(q)^r$. For two random variables $A, B$, $A$ being defined over $\mathcal{A}$, if for a constant $c$, $H_2(A|B=b) > c$, then

$$H(\mathbf{G}(A)|\mathbf{G}, B=b) > r\log_2 q - \frac{2^{r\log_2 q - c}}{\ln 2} \tag{23}$$

With these preparations, we are now ready to prove Theorem 2:

*Proof of Theorem 2:* Define $a \oplus b$ as $a+b \bmod \Lambda_c$. Then for the distribution for $t_i^N, i=1,2$ stated in Theorem 2, $t_1^N \oplus t_2^N$ is independent from $t_1^N$. Therefore we have:

$$H_2\left(t_1^N | t_1^N \oplus t_2^N = t^N\right) = H_2\left(t_1^N\right) = N\log_2 q \tag{24}$$

Let $T$ be the integer defined in Theorem 3. Then according to [36, P 106, Theorem 5.2] [25, Lemma 3], for a given integer $a$, $1 \leq a \leq 2^N$ and $t^N \in \Lambda \cap \mathcal{V}(\Lambda_c)$, with probability $1 - 2^{-(s/2-1)}$:

$$H_2\left(t_1^N | t_1^N \oplus t_2^N = t^N, T = a\right) \geq H_2\left(t_1^N | t_1^N \oplus t_2^N = t^N\right) - \log_2 |T| - s \tag{25}$$

$$= N(\log_2 q - 1) - s \tag{26}$$

In Lemma 1, we have shown that if $\Lambda_c = q\Lambda$, with $q$ being prime, then $\Lambda \cap \mathcal{V}(\Lambda_c)$ is isomorphic to $\mathcal{GF}(q^N)$. The isomorphism is with respect to the addition operation defined in these two sets. Since $t_1^N \in \Lambda \cap \mathcal{V}(\Lambda_c)$, we can write $t_1^N \in \mathcal{GF}(q^N)$. Moreover, since $\mathcal{GF}(q^N)$ is isomorphic to $\mathcal{GF}(q)^N$ in terms of the addition operation defined in these two sets, we can further write $t_1^N \in \mathcal{GF}(q)^N$. Let **G** be taken from the set of linear mappings from $\mathcal{GF}(q)^N$ to $\mathcal{GF}(q)^r$ according to a uniform distribution. Then $\mathbf{G}(t_1^N)$ is well defined.

According to Lemma 3, **G** is a universal hash function. Hence, according to Theorem 4, we have:

$$H\left(\mathbf{G}\left(t_1^N\right) | \mathbf{G}, t_1^N \oplus t_2^N = t^N, T = a\right) \geq r\log_2 q - \frac{2^{r\log_2 q - c}}{\ln 2} \tag{27}$$

where $c$ is given by (26):

$$c = N(\log_2 q - 1) - s \tag{28}$$

Since depending on the value of $t^N$ and $a$ equation (26) holds with probability $1 - 2^{-(s/2-1)}$, from (27), we have

$$H\left(\mathbf{G}\left(t_1^N\right) | \mathbf{G}, t_1^N \oplus t_2^N, T\right) \geq \left(1 - 2^{-(s/2-1)}\right)\left(r\log_2 q - \frac{2^{r\log_2 q - c}}{\ln 2}\right) \tag{29}$$

Note that

$$H(\mathbf{G}(t_1^N)|\mathbf{G}) \leq r \log_2 q \tag{30}$$

Hence in order for $I(\mathbf{G}(t_1^N); t_1^N \oplus t_2^N, T|\mathbf{G})$ to be negligible, we expect $2^{-(s/2-1)}$ and $2^{r \log_q -c}$ to decrease exponentially with respect to $N$. To achieve this, we choose $s = \varepsilon' N$, where $0 < \varepsilon' < \log_2 q - 1$ so that $c$ in (28) is positive. We choose $r$ such that for $\delta > 0$:

$$r \log_2 q < c - N\delta \tag{31}$$
$$= N(\log_2 q - 1) - s - N\delta \tag{32}$$
$$= N(\log_2 q - 1 - \varepsilon' - \delta) \tag{33}$$

We observe that if (31)-(33) are satisfied, $2^{r \log_q -c}$ to decrease exponentially with respect to $N$. We also observe that if we let $\varepsilon = \varepsilon' + \delta$, then (31)-(33) lead to (18).

For these choices of $r$ and $s$, from (29) and (30), we observe that there exists $\beta > 0$, such that

$$I\left(\mathbf{G}\left(t_1^N\right); t_1^N \oplus t_2^N, T|\mathbf{G}\right) \leq e^{-\beta N} \tag{34}$$

We next use the fact that for sufficiently large $N$, most $\mathbf{G}$s have full row rank as shown in Lemma 2. Therefore, for a uniform distribution for $t_i^N, i = 1, 2$, $t_1^N$ and $t_2^N$ being independent, there must exists a $\mathbf{G} = \mathbf{g}$, such that

1) $\mathbf{g}$ has full rank.
2) From Markov inequality,

$$I\left(\mathbf{G}\left(t_1^N\right); t_1^N \oplus t_2^N, T|\mathbf{G} = \mathbf{g}\right) \leq 2e^{-\beta N} \tag{35}$$

Finally, we use Theorem 3, which says $t_1^N \oplus t_2^N, T$ in (35) can be replaced by $\bar{Y}_r^N$. Hence we have proved Theorem 2. ∎

The secrecy generation scheme described above will not be useful if the generated random variable, $\mathbf{g}(t_1^N)$, can not serve as the random seed, $x$, in the AMD tuple as described in Section III. Hence we need the following lemma on the distribution of $\mathbf{g}(t_1^N)$.

*Lemma 4:* If $t_1^N$ is uniformly distributed over $\mathcal{GF}(q^N)$, and $\mathbf{g}$ has full row rank, Then $\mathbf{g}(t_1^N)$ is uniformly distributed over $\mathcal{GF}(q^r)$.

*Proof:* Since **g** has full row rank, and its elements are taken from the field $\mathcal{GF}(q)$, it can always be represented as

$$\mathbf{g} = [\mathbf{I}, \mathbf{P}]\mathbf{O} \tag{36}$$

where **O** is an $N \times N$ invertible matrix. Hence $\mathbf{O}(t_1^N)$ is uniformly distributed over $\mathcal{GF}(q^N)$. **I** is an $r \times r$ identity matrix. Since the sum of any two independent field elements will be uniformly distributed if one of the field element is uniformly distributed, it can be verified that $\mathbf{g}(t_1^N)$ is uniformly distributed over $\mathcal{GF}(q^r)$. ∎

*2) The General Case:* When $(\Lambda, \Lambda_c)$ does not have the self-similar relationship as described in Section V-A1, we can still extract a strongly secure random variable from a lattice point using the same method as shown in Section V-A1. The only difference is that the map between the extracted random variable and the lattice point will not be linear.

Consider a general $N$ dimensional nested lattice codebook $\Lambda \cap \mathcal{V}(\Lambda_c)$. Recall that $R_0$, as defined in (12), is the rate of the codebook. Assume $R_0 > 1$. Let $\lfloor x \rfloor$ be the operation that rounds $x$ to the nearest integer less than or equal to $x$. Define $N_0$ as

$$N_0 = \lfloor \log_2 |\Lambda \cap \mathcal{V}(\Lambda_c)| \rfloor \tag{37}$$

Then

$$N_0 \geq NR_0 - 1 \tag{38}$$

Choose the subset $K$ of the codebook $(\Lambda + d_1^N) \cap \mathcal{V}(\Lambda_c)$ that yields the minimal average decoding error probability with the lattice decoder and has size $|K| = 2^{N_0}$. Define $v$ as the one-to-one mapping from $K$ to $\mathcal{GF}(2^{N_0})$. Then we have the following theorem:

*Theorem 5:* Let $\varepsilon > 0$ be a constant such that

$$R_0 - 1 - \varepsilon > 0 \tag{39}$$

Then for an integer $r_0$, such that

$$0 \leq r_0 \leq N(R_0 - 1 - \varepsilon) \tag{40}$$

there exists a linear mapping **g** from $\mathcal{GF}(2)^{N_0}$ to $\mathcal{GF}(2)^{r_0}$ such that

1) **g** has full row rank $r_0$.

2) When $t_1^N$ is uniformly distributed over $K$, $t_2^N$ is uniformly distributed over $(\Lambda+d_2^N) \cap \mathcal{V}(\Lambda_c)$, $t_1^N, t_2^N$ are independent of each other, we have

$$I\left(\mathbf{g}\left(v(t_1^N)\right); \bar{Y}_r^N\right) \leq 2e^{-\beta N} \tag{41}$$

for a certain $\beta > 0$.

*Proof:* The proof is similar to that of Theorem 2, and is given in Appendix B. ∎

*3) Encoder Construction:* Although both Theorem 2 and Theorem 5 can be used to prove the existence of an encoder with rate arbitrarily close to $\max\{R_0 - 1, 0\}$, with $R_0$ defined in (12), only Theorem 5 is used in the sequel to transmit confidential messages. Theorem 2 is only used to generate strongly secure random seeds, for which Theorem 2 is sufficient by itself. Hence in this section, we discuss Theorem 5 only. The argument we use is as follows:

For a given $\mathbf{g}$ that has full row rank, let $\mathbf{g}'$ be $(N_0 - r_0) \times N_0$ matrix such that $\begin{bmatrix} \mathbf{g}' \\ \mathbf{g} \end{bmatrix}$ is a square matrix that is invertible. Define $\mathbf{S}$ and $\mathbf{S}'$ such that

$$\begin{bmatrix} \mathbf{g}'_{(N_0-r_0) \times N_0} \\ \mathbf{g}_{r_0 \times N_0} \end{bmatrix} v(t_1^N) = \begin{bmatrix} \mathbf{S}'_{(N_0-r_0) \times 1} \\ \mathbf{S}_{r_0 \times 1} \end{bmatrix} \tag{42}$$

Then $\mathbf{S} = \mathbf{g}(v(t_1^N))$. Define $\mathbf{A}$ as the inverse of $\begin{bmatrix} \mathbf{g}' \\ \mathbf{g} \end{bmatrix}$, then the encoder is given by:

$$t_1^N = v^{-1} \mathbf{A} \begin{bmatrix} \mathbf{S}'_{(N_0-r_0) \times 1} \\ \mathbf{S}_{r_0 \times 1} \end{bmatrix} \tag{43}$$

where $\mathbf{S} \in \mathcal{GF}(2^{r_0})$ be the input to the encoder. We assume $\mathbf{S}$ is uniformly distributed over $\mathcal{GF}(2^{r_0})$. $t_1^N \in \Lambda \cap \mathcal{V}(\Lambda_c)$ is the output of the encoder. $\mathbf{S}'$ represents the randomness in the encoding scheme. We observe that, if $\{\mathbf{S}'_{(N_0-r_0) \times 1}, \mathbf{S}_{r_0 \times 1}\}$ is uniformly distributed over $\mathcal{GF}(2)^{N_0}$ and (43) is used as the encoder, $t_1^N$ is also uniformly distributed over the set $K$. Since $\mathbf{G} = \mathbf{g}$ is chosen when $t_1^N$ has a uniform distribution over $K$, this means that when (43) is used as an encoder, the secrecy constraint in Theorem 5, (41), still holds.

Since the encoder (43) uses $N$ channel uses to transmit a $r_0 \times 1$ binary vector, the achieved secrecy rate is

$$R_e = [R_0 - 1 - \varepsilon]^+ \tag{44}$$

where $[x]^+$ equals $x$ if $x \geq 0$ or $0$ otherwise. According to (13), this means $R_e$ can be arbitrarily close to

$$\left[\frac{1}{2}\log_2(\frac{1}{2}+P) - 1\right]^+ \tag{45}$$

*B. Comparison with Other Wiretap Coding Schemes*

Although this work leverages the same technique, namely, privacy amplification as [25], it is distinct from [25] in the following aspects:

Reference [25] proposed that one can invoke any weakly secure scheme multiple times and extract a strongly secure key using privacy amplification. Let $\Theta(x)$ denote the set of functions $ax + b, a > 0, b \neq 0$, and $a, b$ are constants. In our model, each invocation of the weakly secure scheme involves $\Theta(N)$ channel uses, where $N$ is the dimension of the lattice code. Suppose this scheme is invoked for $M$ times. Then the total number of channel uses is $MN$. Let $K$ denote the generated key and $Y_r^{MN}$ be the signals observed by the eavesdropper, then the result in [25] implies [1]

$$\lim_{M \to \infty} I\left(K; Y_r^{MN}\right) = 0 \tag{46}$$

In this work, $\mathbf{g}(t_1^N)$ in Theorem 2 can be viewed as the strongly secure key. Based on Theorem 2, we have

$$\lim_{N \to \infty} -\frac{1}{N}\log_2 I\left(K; Y_r^N\right) > 0 \tag{47}$$

Comparing (47) to (46), we observe (47) is stronger. This is because the strongly secure scheme in Section V-A leverages results specific to nested lattice code, namely Theorem 3 and extracts the key from a single lattice point instead of a sequence of lattice points. Hence, while the scheme we proposed in Section V-A is not as generally applicable as [25] does, we observe that it performs better than applying [25] directly to our model.

---

[1] To simplify the argument, we have omitted several details from [25] including "error reconciliation". Interested readers can refer to [25] for further details.

## VI. BYZANTINE DETECTION

In this section, we describe how to transmit the AMD code using the strong secrecy scheme proposed in Section V and analyze its performance.

To transmit $\{x,h\}$, we use the idea of "message authentication codes with key manipulation security" in [21, Section 4]. Note that for a given $s$, the distribution of hash tag $h$ is in general not uniform. Hence the distribution of $h$ depends on the distribution of $s$. However, if we want to use the strongly secure scheme in Section V-A to transmit $h$ and desire to fix the hash function $\mathbf{G} = \mathbf{g}$, we need to know the distribution of $h$ beforehand, which is difficult since the distribution of $s$ is hard to determine beforehand. To solve this problem, we introduce another random seed $k$ from $\mathcal{GF}(q^r)$, which can be generated via the linear coding scheme in Section V-A. From Lemma 4, $k$ is uniformly distributed over $\mathcal{GF}(q^r)$. Hence $h$ can be transmitted by using $k$ as a one time pad.

The transmission is hence divided into 4 stages:

1) $x \in \mathcal{GF}(q^r)$ is extracted from an $N$ dimensional lattice code as shown in Section V-A1.
2) $k \in \mathcal{GF}(q^r)$ is extracted from an $N$ dimensional lattice code as shown in Section V-A1. Let $\hat{k}$ be the estimate of it computed by node 2. Let $P_1$ be the average power per channel use of the $N$ dimensional lattice code.
3) $u = h \oplus k$ is transmitted by node 1 via the conventional two-hop protocol using $r$-dimensional lattice code with $\log_2 q$ per channel use. In this stage, node 2 remains silent. Let $\hat{u}$ be the estimate of it computed by node 2. Let $P_2$ be the power per channel use of the $r$ dimensional lattice code.
4) $s$ is transmitted via the encoder described in Section V-A2 with $P = \bar{P}(1 - \varepsilon_P)$. $\varepsilon_P$ is a positive constant that can be made arbitrarily small. Let $\hat{s}$ be the estimate of $s$ computed by node 2, which corresponds to $s'$ in Theorem 1.

*Remark 6:* Note that both $P_1$ and $P_2$ are only functions of the rate of their respective lattice code, which is $\log_2 q$. Hence $P_1$ and $P_2$ are only functions of $q$. Therefore, we can increase $r$, while leaving $P_1, P_2$ unchanged. □

We next derive the following important lemma which implies the condition of AMD code stated in Theorem 1 can be fulfilled using the transmission scheme described above:

*Lemma 5:* Let $s_0$ be any $d \times 1$ vector on $\mathcal{GF}(q^r)$. Then

$$I(x; \Delta_x, \Delta_h, \hat{s}|s = s_0) < 4\exp(-\beta N) \tag{48}$$

where $\beta$ is a positive number defined in Theorem 2.

*Proof:* The proof of Lemma 5 is based on the strong secrecy offered by Theorem 2 and Theorem 5, and is provided in Appendix C. ∎

*Remark 7:* Lemma 5 implies that

$$I(x; \Delta_x, \Delta_h, \hat{s}|s) < 4\exp(-\beta N) \tag{49}$$

Since $I(x; s) = 0$, this means

$$I(x; \Delta_x, \Delta_h, \hat{s}, s) < 4\exp(-\beta N) \tag{50}$$

□

*Remark 8:* Note that $I(x; \Delta_x, \Delta_h, \hat{s}|s = s_0)$ does not dependent on the error exponents of the lattice decoder. Also, it does not depend on whether $s_0$ is known by the attacker beforehand. □

We next link Lemma 5 and Theorem 1 with Pinsker's inequality which leads to the following **main result** of this paper:

*Theorem 6:* For the Gaussian two-hop network, for a rate smaller but arbitrarily close to $0.5R_e$ given by (45), and a total number of channel uses $2n = \Theta(N)$:

1) When the relay is honest, the confidential message $W$ can be transmitted at this rate such that all the three terms $\Pr(W \neq \hat{W})$, $I(W; Y_r^n)$ and

$$\Pr\left(\hat{W} \text{ is not accepted by Node 2}|W = \hat{W}\right) \tag{51}$$

decrease exponentially fast with $N$.

2) When the relay is not honest, the probability that the Byzantine attack goes undetected, i.e., the probability that the adversary wins, denoted as $\Pr(A\ wins)$ in (7), decreases exponentially fast with $N$.

*Proof:* We use "HRH" for "hash rule holds" when for $s \neq s'$,

$$x^{d+2} + \sum_{i=1}^{d} s_i x^i = x'^{d+2} + \sum_{i=1}^{d} s'_i x'^i + \Delta_h \tag{52}$$

This means the message $s', x', h'$ will be accepted by node 2. Hence the probability that the adversary wins is given by:

$$\Pr(A \ wins) = \sum_{\substack{x, \Delta_x \\ \Delta_h, s' \neq s_0}} \Pr(\text{HRH}|x, \Delta_h, \Delta_x, s = s_0, s') \Pr(x|\Delta_h, \Delta_x, s = s_0, s') \Pr(\Delta_h, \Delta_x, s'|s = s_0) \tag{53}$$

Define $Q(A \ wins)$ as the term (53) with $\Pr(x|\Delta_h, \Delta_x, s = s_0, s')$ replaced by $\Pr(x)$.

$$Q(A \ wins) = \sum_{\substack{x, \Delta_x \\ \Delta_h, s' \neq s_0}} \Pr(\text{HRH}|x, \Delta_h, \Delta_x, s = s_0, s') \Pr(x) \Pr(\Delta_h, \Delta_x, s'|s = s_0) \tag{54}$$

Note that $Q(A \ wins)$ would be the probability that the Byzantine adversary wins if $x$ and $\Delta_h, \Delta_x, s, s'$ are truly independent. To evaluate the effect of being otherwise, we next bound the difference between $\Pr(A \ wins)$ and $Q(A \ wins)$.

$$|\Pr(A \ wins) - Q(A \ wins)| \tag{55}$$

$$\leq \sum_{\substack{x, \Delta_x \\ \Delta_h, s' \neq s_0}} \Pr(\text{HRH}|x, \Delta_h, \Delta_x, s = s_0, s') |\Pr(x|\Delta_h, \Delta_x, s = s_0, s') - \Pr(x)| \Pr(\Delta_h, \Delta_x, s'|s = s_0) \tag{56}$$

$$\leq \sum_{\substack{x, \Delta_x \\ \Delta_h, s' \neq s_0}} |\Pr(x|\Delta_h, \Delta_x, s = s_0, s') - \Pr(x)| \Pr(\Delta_h, \Delta_x, s'|s = s_0) \tag{57}$$

$$= \sum_{\substack{x, \Delta_x \\ \Delta_h, s' \neq s_0}} |\Pr(x|\Delta_h, \Delta_x, s', s = s_0) - \Pr(x|s = s_0)| \Pr(\Delta_h, \Delta_x, s'|s = s_0) \tag{58}$$

$$= \sum_{\substack{x, \Delta_x \\ \Delta_h, s' \neq s_0}} |\Pr(x, \Delta_h, \Delta_x, s'|s = s_0) - \Pr(x|s = s_0) \Pr(\Delta_h, \Delta_x, s'|s = s_0)| \tag{59}$$

$$\leq \sum_{\substack{x, \Delta_x \\ \Delta_h, s'}} |\Pr(x, \Delta_h, \Delta_x, s'|s = s_0) - \Pr(x|s = s_0) \Pr(\Delta_h, \Delta_x, s'|s = s_0)| \tag{60}$$

Then we use Pinsker's inequality [37, Theorem 2.33]:

$$I(A; B) \geq \frac{1}{2 \ln 2} D^2(p(A, B), p(A) p(B)) \tag{61}$$

where $D(p(x), q(x)) = \sum_x |p(x) - q(x)|$.

Let $p(A)$ be $\Pr(x|s = s_0)$. Let $p(B)$ be $\Pr(\Delta_h, \Delta_x, s'|s = s_0)$. Let $p(A, B)$ be given by:

$$p(A, B) = \Pr(x, \Delta_h, \Delta_x, s'|s = s_0) \tag{62}$$

Then from Lemma 5, (60) is bounded by $\sqrt{(8\ln 2)\exp(-\beta N)}$ because of Pinsker's inequality. Hence we have:

$$|\Pr(A\ wins) - Q(A\ wins)| \leq \sqrt{(8\ln 2)\exp(-\beta N)} \tag{63}$$

From Theorem 1, $Q(A\ wins)$ is bounded by $\frac{d+1}{q^r}$. Hence

$$\Pr(A\ wins) \leq \sqrt{(8\ln 2)\exp(-\beta N)} + \frac{d+1}{q^r} \tag{64}$$

Each $\{s\}$ conveys $dr\log_2 q$ bits of information, where $r$ is defined in Theorem 2. Recall that the total number of channel uses is denoted by $2n$. The relay node transmits during $n$ channel uses. Node 1 transmits during the other $n$ channel uses. When node 1 transmits, node 2 may or may not transmit depending on which of the 4 stages described at the beginning of this section is being executed. For the four-stage transmission scheme, $n$ is given by:

$$n = 2N + r + \left\lceil \frac{dr\log_2 q}{NR_e} \right\rceil N \tag{65}$$

This is because $N$ channel uses are needed to transmit $x$ or $k$, and $r$ channel uses are needed to transmit $k \oplus h$. The third term in (65) is the number of channel uses needed to transmit $s$, where $\lceil x \rceil$ is the operation that rounds $x$ to the nearest integer greater than or equal to $x$.

The overall secrecy rate $R_T$ is given by

$$R_T = \frac{dr\log_2 q}{2n} \tag{66}$$

From (65), we observe $R_T$ can be made arbitrarily close to $0.5R_e$ by choosing a sufficiently large $d$.

Let $P_T$ denote the transmission power averaged over the channel uses during which a node transmits. Based on the four stage transmission scheme, $P_T$ of node 1 and the relay are the same. $P_T$ of node 2 is smaller since it does not transmit during the third stage. Hence we only need to make sure $P_T$ of node 1 does not exceed the power constraint $P$. $P_T$ of node 1 is given by

$$P_T = \frac{P_1 2N + P_2 r + P\left(\frac{dr\log_2 q}{R_e}\right)}{n} \tag{67}$$

$P_T$ can be made arbitrarily close to but strictly smaller than $\bar{P}$ by choosing a sufficiently large $d$ and a sufficiently small $\varepsilon_P$.

Once $R_T$ and $P_T$ is fixed, $d$ is fixed. On the other hand, as shown by (65) and (18), for a fixed $d$, $n$ increases linearly with respect to $N$.

Select $r$ as in (18) such that $r$ increases linearly with respect to $N$. Then, from (64), we observe that the probability that the adversary wins decreases exponentially fast with $N$. Hence we have the bound on $\Pr(A\ wins)$ stated in the theorem.

We next check whether the secrecy constraint is satisfied:

$$I\left(s; Y_r(i), 0 \leq i \leq 3\right) \tag{68}$$

$$\leq I\left(x; Y_r(0)\right) + I\left(h; Y_r(1), Y_r(2)\right) + I\left(s; Y_r(3)\right) \tag{69}$$

In (69), the first term decreases exponentially fast with respect to $N$ due to Theorem 2. For the second term, we have

$$I\left(h; Y_r(1), Y_r(2)\right) \leq I\left(h; \bar{Y}_r(1), Z_r(1), h \oplus k\right) \tag{70}$$

$$= I\left(h; \bar{Y}_r(1), h \oplus k\right) \tag{71}$$

$$= I\left(h; h \oplus k\right) + I\left(h; \bar{Y}_r(1) | h \oplus k\right) \tag{72}$$

$$= I\left(h; \bar{Y}_r(1) | h \oplus k\right) \tag{73}$$

$$\leq I\left(h, k; \bar{Y}_r(1)\right) = I\left(k; \bar{Y}_r(1)\right) \tag{74}$$

Hence, the second term is bounded by $I(k, \bar{Y}_r(1)$, which also decreases exponentially fast with respect to $N$ due to Theorem 2. The third term decreases exponentially fast with respect to $\frac{dr \log_2 q}{R_e}$ due to Theorem 5. Hence (68) decreases exponentially fast with respect to $N$.

Finally, we check whether the confidential message $W$, which corresponds to $s$ in our scheme, can be transmitted reliably. We observe that the probability $\Pr(W \neq \hat{W})$ does decrease exponentially fast with respect to $N$ because the decoding error probability of the lattice decoder decreases at this speed, as stated in the end of Section IV.

The probability

$$\Pr\left(\hat{W} \text{ is not accepted by Node 2} | W = \hat{W}\right) \tag{75}$$

depends on whether $x, k, k \oplus h$ can be transmitted reliably. Since they are also transmitted with the nested lattice code and decoded with a lattice decoder, the probability of decoding error when transmitting $x, k, k \oplus h$ also decreases exponentially with respect to the dimension of the lattice, which in turn increases linearly with $N$. Hence (75) also decreases exponentially fast with respect to $N$.

Hence we have proved the theorem.

∎

*Remark 9:* It is evident from (63) that if Lemma 5 were weakened to just proving the left-hand side converges to $0$, which is the case if the conventional strong secrecy notion like the one in [27] is used, then it would not be possible to preserve the exponentially decreasing detection property offered by the AMD code. Hence in this problem, the commonly recognized strong secrecy notion is insufficient, and a stronger notion, as described by (19), is required.

## VII. Conclusion

In this work, we developed a coding scheme which provides strong secrecy by combining nested lattice codes and universal hash functions. In our previous work [23], the representation theorem for nested lattice codes is used to bound the Shannon entropy. Here we showed the same theorem is also useful in bounding another information theoretic measure, i.e., the Rényi entropy, which in turn leads to the desired strong secrecy results in a Gaussian setting. We showed that this coding scheme can be used with AMD codes to perform Byzantine detection for a Gaussian two-hop network where the relay is both an eavesdropper and a Byzantine attacker. Using this code, we showed that the probability that a Byzantine adversary wins decreases exponentially fast with respect to the number of channel uses.

It should be noted that, in this work, we have assumed that the channel gains are known by each node before the communication starts. It should be recognized that the Byzantine attacker at the relay node may attempt to manipulate the channel estimation process, for example, by broadcast incorrect pilot signals, to gain an advantage. Detection of this type of misbehavior is closely related to the physical layer implementation of the system and is left as future work.

## Appendix A
## Proof of Lemma 1

When $\Lambda_c = q\Lambda$ and the generation matrix of $\Lambda$ has full rank, there are $q^N$ lattice points in $(\Lambda + d^N) \cap \mathcal{V}(\Lambda_c)$. Each point in $(\Lambda + d^N) \cap \mathcal{V}(\Lambda_c)$ can be represented by its coordinates, which is a vector composed of $N$ integers: $\{c_1, ..., c_N\}$.

We next prove the following mapping is an isomorphism from $(\Lambda + d^N) \cap \mathcal{V}(\Lambda_c)$ to the group of a finite field $\mathcal{GF}(q^N)$:

$$\mathbf{I} : \mathbf{I}(c_1, ... c_N) = \{c_1 \bmod q + (c_2 \bmod q)x ... + (c_N \bmod q)x^{N-1}\} \tag{76}$$

First we prove that two elements in $(\Lambda + d^N) \cap \mathcal{V}(\Lambda_c)$ can not be mapped to the same element in $\mathcal{GF}(q^N)$. This can be proved via contradiction: Suppose they can. Then, we have two points $x$, and $y$, whose coordinates are $\{a_1, ..., a_N\}$ and $\{b_1, ..., b_N\}$ respectively, such that

$$a_i - b_i \bmod q = 0 \quad i = 1, ..., N \tag{77}$$

$$\exists j, \quad a_j \neq b_j \tag{78}$$

This means $x - y \in q\Lambda = \Lambda_c$. Let $z \in \Lambda_c$ be $x - y$. Then $x = y + z$ and $z \neq 0$.

Define the quantization operator $Q_{\Lambda_c}(x)$ as

$$Q_{\Lambda_c}(x) = \arg \min_{t \in \Lambda_c} \|t - x\| \tag{79}$$

where $\|t - x\|$ denotes the Euclidean distance between $t$ and $x$. $Q_{\Lambda_c}(x)$ has the following property: $\forall z \in \Lambda_c$, $Q_{\Lambda_c}(x + z) = Q_{\Lambda_c}(x) + z$. This can be shown as follows:

$$Q_{\Lambda_c}(x + z) = \arg \min_{t \in \Lambda_c} \|t - x - z\| \tag{80}$$

$$= \arg \min_{t - z \in \Lambda_c} \|(t - z) - x\| \tag{81}$$

$$= \arg \min_{t' \in \Lambda_c} \|t' - x\| + z \tag{82}$$

$$= Q_{\Lambda_c}(x) + z \tag{83}$$

Since $x, y \in \mathcal{V}(\Lambda_c)$. This means $Q_{\Lambda_c}(x) = 0$ and $Q_{\Lambda_c}(y) = 0$. However we can also write $Q_{\Lambda_c}(x) = Q_{\Lambda_c}(y + z) = Q_{\Lambda_c}(y) + z = z \neq 0$. This leads to a contradiction.

Since $\mathbf{I}$ cannot map two different lattice points to the same field element, and the set $(\Lambda + d^N) \cap \mathcal{V}(\Lambda_c)$ has the same cardinality as $\mathcal{GF}(q^N)$, $\mathbf{I}$ must be a one-to-one mapping.

Finally, it is easy to verify that $\mathbf{I}$ preserves the addition operation:

$$\mathbf{I}(x + y) = \mathbf{I}(x) + \mathbf{I}(y) \tag{84}$$

This completes the proof that $\mathbf{I}$ is an isomorphism.

# APPENDIX B
## PROOF OF THEOREM 5

For the distribution for $t_i^N, i = 1, 2$ stated in Theorem 5, $t_1^N \oplus t_2^N$ is independent from $t_1^N$. Therefore:

$$H_2\left(t_1^N | t_1^N \oplus t_2^N = t^N\right) = H_2\left(t_1^N\right) = N_0 \qquad (85)$$

Then, as in (26), with probability $1 - 2^{-(s/2-1)}$:

$$H_2\left(t_1^N | t_1^N \oplus t_2^N = t^N, T = a\right) \qquad (86)$$

$$\geq H_2\left(t_1^N | t_1^N \oplus t_2^N = t^N\right) - \log_2 |T| - s = N_0 - N - s \qquad (87)$$

We next use the fact that when $\mathbf{G}$ is uniformly distributed over the set of linear functions from $\mathcal{GF}(2)^{N_0}$ to $\mathcal{GF}(2)^{r_0}$, the following equation holds according to Theorem 4:

$$H\left(\mathbf{G}\left(v(t_1^N)\right) | \mathbf{G}, t_1^N \oplus t_2^N = t^N, T = a\right) \geq r_0 - \frac{2^{r_0-c}}{\ln 2} \qquad (88)$$

where $c = N_0 - N - s$.

Hence

$$H\left(\mathbf{G}\left(v(t_1^N)\right) | \mathbf{G}, t_1^N \oplus t_2^N, T\right) \geq \left(1 - 2^{-(s/2-1)}\right)\left(r_0 - \frac{2^{r_0-c}}{\ln 2}\right) \qquad (89)$$

In order for $2^{-(s/2-1)}$ to decrease exponentially fast with respect to $N$, we choose $s = \varepsilon N$, where $0 < \varepsilon < R_0 - 1$ so that $c$ is positive. Choose $r_0$ such that for $\delta > 0$:

$$r_0 < c - N\delta/2 = N_0 - N - s - N\delta/2 \qquad (90)$$

so that $2^{r_0-c}$ decreases exponentially fast with respect to $N$. Recall by (38), we have $N_0 \geq NR_0 - 1$. Hence a sufficient condition for (90) to hold is to require

$$r_0 < N(R_0 - 1) - s - N\delta \qquad (91)$$

This yields (40). For this $r_0$ and $s$, from (89), we observe that there exists $\beta > 0$, such that

$$I\left(\mathbf{G}\left(v(t_1^N)\right); t_1^N \oplus t_2^N, T | \mathbf{G}\right) \leq e^{-\beta N} \qquad (92)$$

We next use the fact that for sufficiently large $N$, most $\mathbf{G}$ has full row rank as shown in Lemma 2. Therefore, under a uniform distribution for $t_i^N, i = 1, 2$, $t_1^N$ and $t_2^N$ being independent, there must exists a $\mathbf{G} = \mathbf{g}$, such that

1) $\mathbf{g}$ has full rank.
2) $I\left(\mathbf{G}\left(v(t_1^N)\right); t_1^N \oplus t_2^N, T | \mathbf{G} = \mathbf{g}\right) \leq 2e^{-\beta N}$

Hence we have proved Theorem 5.

## APPENDIX C

### PROOF OF LEMMA 5

The following notation is used in the proof: $X_i(j), i = 0, ..., 3, X_r(j)$ denote the signals transmitted by node $1, 2$ and the relay during the $j$th stage, $j = 0, ..., 3$. Similarly, $Y_i(j), i = 1, 2, Y_r(j), Z_r(j), Z_R(j)$ denote the signals and channel noise observed during the $j$th stage. $\hat{X}_r(i), i = 0, ..., 3$ denotes the estimate for $X_r(i)$ computed by node 2. To simplify the notation, we omit the superscript for these signals which were used to indicate their dimensions.

As described in Section VI, the 0th stage is used to transmit $x$. The 1st stage is used to transmit $k$. The 2nd stage is used to transmit $k \oplus h$. The 3rd stage is used to transmit $s$.

We next explain how to upper bound the following quantity:

$$I\left(x; \Delta_x, \Delta_h, \hat{s} | s = s_0\right) \tag{93}$$

Let $\oplus$ in $x \oplus y$ denote the addition operation in the field where $x$ and $y$ are taken from. Let $-x$ denote the element such that $(-x) \oplus x = 0$. Recall that $\mathbf{g}$ is the linear mapping whose existence is proved in Theorem 2. With these notations, we can write $\Delta_x$ as:

$$\Delta_x = \mathbf{g}\left(\hat{X}_r(0) \oplus (-X_2(0))\right) \oplus (-x) \tag{94}$$

$$= \mathbf{g}\left(\hat{X}_r(0) \oplus (-X_2(0))\right) \oplus \mathbf{g}\left(-X_1(0)\right) \tag{95}$$

$$= \mathbf{g}\left(\hat{X}_r(0) \oplus \left(-(X_2(0) \oplus X_1(0))\right)\right) \tag{96}$$

Since $\Delta_x$ is a function of $\hat{X}_r(0)$ and $X_2(0) \oplus X_1(0)$, (93) is upper bounded by:

$$I\left(x; \hat{X}_r(0), X_1(0) \oplus X_2(0), \Delta_h, \hat{s} | s = s_0\right) \tag{97}$$

$\hat{X}_r(0)$ is computed from $Y_2(0)$ by node 2. Hence (97) is upper bounded by:

$$I\left(x; Y_2(0), X_1(0) \oplus X_2(0), \Delta_h, \hat{s} | s = s_0\right) \tag{98}$$

$$\leq I\left(x; X_r(0), Z_R(0), X_1(0) \oplus X_2(0), \Delta_h, \hat{s} | s = s_0\right) \tag{99}$$

$$= I\left(x; X_r(0), X_1(0) \oplus X_2(0), \Delta_h, \hat{s} | s = s_0\right)$$

$$+ I\left(x; Z_R\left(0\right) | X_r\left(0\right), X_1\left(0\right) \oplus X_2\left(0\right), \Delta_h, \hat{s}, s = s_0\right) \quad (100)$$

Recall that $Z_R(0)$ is the noise observed by Node 2 during the stage responsible for transmitting $x$. We observe that it is independent from all the other terms in the second term of (100). This is because $\Delta_h$, $\hat{s}$, $s$ are only related to signals transmitted in later stages. The relay node has no knowledge of $Z_R(0)$. Hence $Z_R(0)$ can not affect the relaying strategy. As a result, (100) equals

$$I\left(x; X_r\left(0\right), X_1\left(0\right) \oplus X_2\left(0\right), \Delta_h, \hat{s} | s = s_0\right) \quad (101)$$

Recall that $M_r$ denotes the randomness available to the relay node. Then, the expression in (101) is upper bounded by

$$I\left(x; M_r, X_r\left(0\right), Y_r\left(0\right), X_1\left(0\right) \oplus X_2\left(0\right), \Delta_h, \hat{s} | s = s_0\right) \quad (102)$$

$$= I\left(x; M_r, Y_r\left(0\right), X_1\left(0\right) \oplus X_2\left(0\right), \Delta_h, \hat{s} | s = s_0\right) +$$

$$I\left(x; X_r\left(0\right) | M_r, Y_r\left(0\right), X_1\left(0\right) \oplus X_2\left(0\right), \Delta_h, \hat{s}, s = s_0\right) \quad (103)$$

Since $X_r(0)$ is computed from $Y_r(0)$ at the relay node, it is a deterministic function of $Y_r(0)$, $M_r$ and potentially $s_0$. Hence the second term in (103) is 0, and (103) equals:

$$I\left(x; M_r, Y_r\left(0\right), X_1\left(0\right) \oplus X_2\left(0\right), \Delta_h, \hat{s} | s = s_0\right) \quad (104)$$

We next examine $\Delta_h$ in (104). Recall that $u$ is defined as $k \oplus h$. $\hat{u}$ and $\hat{k}$ are the estimates for $u$ and $k$ computed by node 2 respectively. With these notations, we can express $\Delta_h$ as:

$$\Delta_h = \hat{u} \oplus (-\hat{k}) \oplus (-h) \quad (105)$$

$$= \hat{u} \oplus ((-k) \oplus (-\Delta_k)) \oplus (-h) \quad (106)$$

$$= \hat{u} \oplus (-(k \oplus h)) \oplus (-\Delta_k) \quad (107)$$

As seen from (105)-(107), $\Delta_h$ is a function of $\hat{u}$, $k \oplus h$, and $\Delta_k$. Therefore (104) can be upper bounded by:

$$I\left(x; M_r, Y_r\left(0\right), X_1\left(0\right) \oplus X_2\left(0\right), \hat{u}, k \oplus h, \Delta_k, \hat{s} | s = s_0\right) \quad (108)$$

Note that $\hat{u}$ is computed from $Y_2(2)$ by node 2. Therefore (108) is upper bounded by:

$$I\left(x; M_r, Y_r\left(0\right), X_1\left(0\right) \oplus X_2\left(0\right), Y_2\left(2\right), k \oplus h, \Delta_k, \hat{s} | s = s_0\right) \quad (109)$$

$$\leq I(x; M_r, Y_r\left(0\right), X_1\left(0\right) \oplus X_2\left(0\right), X_r\left(2\right), Z_R\left(2\right), k \oplus h, \Delta_k, \hat{s} | s = s_0) \quad (110)$$

$$=I(x; M_r, Y_r(0), X_1(0) \oplus X_2(0), X_r(2), k \oplus h, \Delta_k, \hat{s}|s = s_0)$$
$$+ I(x; Z_R(2)|M_r, Y_r(0), X_1(0) \oplus X_2(0), X_r(2), k \oplus h, \Delta_k, \hat{s}, s = s_0) \quad (111)$$

Again $Z_R(2)$ is independent from all the other terms in the second term of (111). Hence (111) equals:

$$I(x; M_r, Y_r(0), X_1(0) \oplus X_2(0), X_r(2), k \oplus h, \Delta_k, \hat{s}|s = s_0) \quad (112)$$

For $\Delta_k$, we have:

$$\Delta_k = \mathbf{g}\left(\hat{X}_r(1) \oplus (-X_2(1))\right) \oplus (-k) \quad (113)$$
$$= \mathbf{g}\left(\hat{X}_r(1) \oplus (-X_2(1))\right) \oplus \mathbf{g}(-X_1(1)) \quad (114)$$
$$= \mathbf{g}\left(\hat{X}_r(1) \oplus (-(X_2(1) \oplus X_1(1)))\right) \quad (115)$$

Hence $\Delta_k$ is a function of $\hat{X}_r(1)$, $X_2(1) \oplus X_1(1)$. Therefore (112) can be upper bounded by:

$$I(x; M_r, Y_r(0), X_1(0) \oplus X_2(0), Y_r(2), k \oplus h, \hat{X}_r(1), X_1(1) \oplus X_2(1), \hat{s}|s = s_0) \quad (116)$$

$\hat{X}_r(1)$ is computed from $Y_2(1)$ by node 2. Hence (116) is upper bounded by:

$$I(x; M_r, Y_r(0), X_1(0) \oplus X_2(0), Y_r(2), k \oplus h, Y_2(1), X_1(1) \oplus X_2(1), \hat{s}|s = s_0) \quad (117)$$
$$\leq I(x; M_r, Y_r(0), X_1(0) \oplus X_2(0), Y_r(2), k \oplus h, X_r(1), Z_R(1),$$
$$X_1(1) \oplus X_2(1), \hat{s}|s = s_0) \quad (118)$$
$$= I(x; M_r, Y_r(0), X_1(0) \oplus X_2(0), Y_r(2), k \oplus h, X_r(1), X_1(1) \oplus X_2(1), \hat{s}|s = s_0)$$
$$+ I(x; Z_R(1)|M_r, Y_r(0), X_1(0) \oplus X_2(0), Y_r(2),$$
$$k \oplus h, X_r(1), X_1(1) \oplus X_2(1), \hat{s}, s = s_0) \quad (119)$$
$$= I(x; M_r, Y_r(0), X_1(0) \oplus X_2(0), Y_r(2), k \oplus h, Y_r(1), X_1(1) \oplus X_2(1), \hat{s}|s = s_0) \quad (120)$$

Finally, $\hat{s}$ is computed from $Y_2(3), X_2(3)$ by node 2. Hence (120) is upper bounded by:

$$I(x; M_r, Y_r(0), X_1(0) \oplus X_2(0), Y_r(2), k \oplus h, Y_r(1),$$
$$X_1(1) \oplus X_2(1), Y_2(3), X_2(3)|s = s_0) \quad (121)$$
$$\leq I(x; M_r, Y_r(0), X_1(0) \oplus X_2(0), Y_r(2), k \oplus h, Y_r(1),$$
$$X_1(1) \oplus X_2(1), X_r(3), X_2(3)|s = s_0) \quad (122)$$

Since $X_r(3)$ is a deterministic function of $M_r$, $Y_r(3)$ and potentially $s_0$, we can upper bound (122) with the following term by replacing $X_r(3)$ with $Y_r(3)$:

$$I(x; M_r, Y_r(0), X_1(0) \oplus X_2(0), Y_r(2), k \oplus h, Y_r(1), X_1(1) \oplus X_2(1),$$
$$Y_r(3), X_2(3) | s = s_0) \tag{123}$$
$$\leq I(x; M_r, Y_r(0), X_1(0) \oplus X_2(0), Y_r(2), k \oplus h, Y_r(1), X_1(1) \oplus X_2(1),$$
$$X_1(3), Z_r(3), X_2(3) | s = s_0) \tag{124}$$

Equation (124) follows from $Y_r(3) = X_1(3) + X_2(3) + Z_r(3)$. We then use the fact that the stochastic encoder used by node 1 to transmit $s$ is independent from the stochastic mapping used at other stages. Hence, we have:

$$I(x; X_1(3), X_2(3), Z_r(3) | M_r, Y_r(0), X_1(0) \oplus X_2(0), \tag{125}$$
$$Y_r(2), k \oplus h, Y_r(1), X_1(1) \oplus X_2(1), s = s_0) = 0 \tag{126}$$

and (124) equals:

$$I(x; M_r, Y_r(0), X_1(0) \oplus X_2(0), Y_r(2), k \oplus h, Y_r(1), X_1(1) \oplus X_2(1) | s = s_0) \tag{127}$$
$$= I(x; M_r | s = s_0)$$
$$+ I(x; Y_r(0), X_1(0) \oplus X_2(0), Y_r(2), k \oplus h, Y_r(1), X_1(1) \oplus X_2(1) | M_r, s = s_0) \tag{128}$$

Next we note that since $I(x; M_r | s = s_0) = 0$, (128) equals:

$$I(x; Y_r(0), X_1(0) \oplus X_2(0), Y_r(2), k \oplus h, Y_r(1), X_1(1) \oplus X_2(1) | M_r, s = s_0) \tag{129}$$

Equation (129) is upper bounded by:

$$I(x, h; Y_r(0), X_1(0) \oplus X_2(0), Y_r(2), k \oplus h, Y_r(1), X_1(1) \oplus X_2(1) | M_r, s = s_0) \tag{130}$$

Recall that the notation $\bar{Y}_r$, as introduced in (16), denotes the quantity obtained by subtracting the channel noise $N_r$ from $Y_r$. Following this notation, we can upper bound (130) as:

$$I(x, h; \bar{Y}_r(0), Z_r(0), X_1(0) \oplus X_2(0),$$
$$\bar{Y}_r(2), Z_r(2), k \oplus h, \bar{Y}_r(1), Z_r(1), X_1(1) \oplus X_2(1) | M_r, s = s_0) \tag{131}$$
$$= I(x, h; \bar{Y}_r(0), X_1(0) \oplus X_2(0),$$

$$\bar{Y}_r(2), k \oplus h, \bar{Y}_r(1), X_1(1) \oplus X_2(1) | M_r, Z_r(i), i = 1, 2, 3, s = s_0) \tag{132}$$

which is further upper bounded by:

$$H\left(\bar{Y}_r(0), X_1(0) \oplus X_2(0) | M_r, Z_r(i), i = 1, 2, 3, s = s_0\right)$$
$$+ H\left(\bar{Y}_r(2), k \oplus h, \bar{Y}_r(1), X_1(1) \oplus X_2(1) | M_r, Z_r(i), i = 1, 2, 3, s = s_0\right)$$
$$- H(\bar{Y}_r(0), X_1(0) \oplus X_2(0), \bar{Y}_r(2), k \oplus h, \bar{Y}_r(1), X_1(1) \oplus X_2(1) |$$
$$x, h, M_r, Z_r(i), i = 1, 2, 3, s = s_0) \tag{133}$$
$$= H\left(\bar{Y}_r(0), X_1(0) \oplus X_2(0) | M_r, Z_r(i), i = 1, 2, 3, s = s_0\right)$$
$$+ H\left(\bar{Y}_r(2), k \oplus h, \bar{Y}_r(1), X_1(1) \oplus X_2(1) | M_r, Z_r(i), i = 1, 2, 3, s = s_0\right)$$
$$- H(\bar{Y}_r(0), X_1(0) \oplus X_2(0) | x, h, M_r, Z_r(i), i = 1, 2, 3, s = s_0)$$
$$- H(\bar{Y}_r(2), k \oplus h, \bar{Y}_r(1), X_1(1) \oplus X_2(1) | \bar{Y}_r(0), X_1(0) \oplus X_2(0),$$
$$x, h, M_r, Z_r(i), i = 1, 2, 3, s = s_0) \tag{134}$$

We then use the two Markov chains shown below:

$$\{\bar{Y}_r(0), X_1(0) \oplus X_2(0)\} - \{x, M_r, Z_r(i), i = 1, 2, 3, s\} - h \tag{135}$$

$$\{\bar{Y}_r(2), k \oplus h, \bar{Y}_r(1), X_1(1) \oplus X_2(1)\} - \{h, M_r, Z_r(i), i = 1, 2, 3, s\}$$
$$- \{x, \bar{Y}_r(0), X_1(0) \oplus X_2(0)\} \tag{136}$$

The Markov relation in (135) holds because given $x$, the distribution of $\{\bar{Y}_r(0), X_1(0) \oplus X_2(0)\}$ only depends on the randomness in the transmitter of node 1 and 2 during stage 0. The Markov chain in (135) follows because:

$$k \oplus h - \{h, M_r, Z_r(i), i = 1, 2, 3, s\} - \{x, \bar{Y}_r(0), X_1(0) \oplus X_2(0)\} \tag{137}$$

and

$$\{\bar{Y}_r(2), \bar{Y}_r(1), X_1(1) \oplus X_2(1)\} - \{k \oplus h, h, M_r, Z_r(i), i = 1, 2, 3, s\}$$
$$- \{x, \bar{Y}_r(0), X_1(0) \oplus X_2(0)\} \tag{138}$$

are Markov chains. Equation (137) is a Markov chain, because, given $h$, the distribution of $k \oplus h$ only depends on $k$, which is independent from all the remaining terms in (137). Equation (138) is a Markov chain, because, given $k \oplus h$ and $h$, which implies $k$ is given, the distribution of $\{\bar{Y}_r(2), \bar{Y}_r(1), X_1(1) \oplus X_2(1)\}$ only depends on the randomness in the transmitter of node 1 and 2 during stage 1 and stage 2.

Applying the two Markov chains (135) and (136) to the last two terms in (134), we find that it equals:

$$I\left(x; \bar{Y}_r(0), X_1(0) \oplus X_2(0) | M_r, Z_r(i), i=1,2,3, s=s_0\right)$$
$$+ I\left(h; \bar{Y}_r(2), k \oplus h, \bar{Y}_r(1), X_1(1) \oplus X_2(1) | M_r, Z_r(i), i=1,2,3, s=s_0\right) \quad (139)$$

For the first term in (139), since $X_1(0) \oplus X_2(0) = \bar{Y}_r(0) \mod \Lambda_c$ and hence is a function of $\bar{Y}_r(0)$, we have

$$I\left(x; \bar{Y}_r(0), X_1(0) \oplus X_2(0) | M_r, Z_r(i), i=1,2,3, s=s_0\right) \quad (140)$$
$$= I\left(x; \bar{Y}_r(0) | M_r, Z_r(i), i=1,2,3, s=s_0\right) = I\left(x; \bar{Y}_r(0)\right) \quad (141)$$

Since $x$ is extracted from a lattice point in $\mathcal{GF}(q^N)$ based on the strong secrecy scheme described in Section V-A1, from Theorem 2, we have $I\left(x; \bar{Y}_r(0)\right) < 2\exp(-\beta N)$.

For the second term in (139), note that $\bar{Y}_r(2)$ is just $X_1(2)$, because node 2 remains silent at this stage. Therefore, this term can be expressed as:

$$I\left(h; X_1(2), k \oplus h, \bar{Y}_r(1), X_1(1) \oplus X_2(1) | M_r, Z_r(i), i=1,2,3, s=s_0\right) \quad (142)$$
$$= I\left(h; \bar{Y}_r(1), X_1(2) | M_r, Z_r(i), i=1,2,3, s=s_0\right)$$
$$+ I\left(h; k \oplus h, X_1(1) \oplus X_2(1) | \bar{Y}_r(1), X_1(2), M_r, Z_r(i), i=1,2,3, s=s_0\right) \quad (143)$$

The second term in (143) is 0 since $k \oplus h$ is a deterministic function of $X_1(2)$ and $X_1(1) \oplus X_2(1)$ is a deterministic function of $\bar{Y}_r(1)$. Therefore (143) equals

$$I\left(h; \bar{Y}_r(1), X_1(2) | M_r, Z_r(i), i=1,2,3, s=s_0\right) \quad (144)$$
$$= I\left(h; \bar{Y}_r(1), X_1(2)\right) \quad (145)$$

Since $X_1(2)$ is determined by $h \oplus k$, (145) is upper bounded by:

$$I\left(h; \bar{Y}_r(1), h \oplus k\right) \quad (146)$$
$$= I(h; h \oplus k) + I\left(h; \bar{Y}_r(1) | h \oplus k\right) \quad (147)$$
$$= I\left(h; \bar{Y}_r(1) | h \oplus k\right) \quad (148)$$

$$\leq I\left(h, k; \bar{Y}_r(1)\right) = I\left(k; \bar{Y}_r(1)\right) \tag{149}$$

Since $k$ is extracted from a lattice point in $\mathcal{GF}(q^N)$ based on the strong secrecy scheme described in Section V-A1, hence from Theorem 2, (149) is bounded by $2\exp(-\beta N)$.

Therefore (139) is bounded by $4\exp(-\beta N)$. Hence we have Lemma 5.